\documentstyle[hip-artc]{article}  
\volnumber{9}  \edyear{1999}  \frompage{000} \topage{000}                %| 
\recrevdate{1 January 1999}                                              %| 
\def\rightmark{Muon-induced fission} 
\def\leftmark{Volker E. Oberacker}

\newcommand{\lambdabar}{{\mathchar'26\mkern-9mu\lambda}}
 
\title{Prompt muon-induced fission: a probe for nuclear energy dissipation} 
\authors{
{\twerm Volker E. Oberacker %
}\\[2.812mm]
{\normalsize
\hspace*{-8pt} Department of Physics \& Astronomy, Vanderbilt University, \\ 
Nashville, TN 37235, USA, E-mail: volker.e.oberacker@vanderbilt.edu\\[0.2ex] 
}}
 
\abstract{We solve the time-dependent Dirac equation for a muon which
is initially bound to a fissioning actinide nucleus. The computations
are carried out on a 3-D cartesian lattice utilizing the Basis-Spline
collocation method. The muon dynamics is sensitive to the nuclear energy
dissipation between the outer fission barrier and the scission point.
From a comparison with experimental data we find a dissipated
energy of about $10$ MeV and a fission time delay due to friction of 
order $2 \times 10^{-21}$ s.}
 
\begin{document}
 
\maketitle

\section{Introduction}
There are two different mechanisms that contribute to nuclear energy
dissipation, i.e. the irreversible transfer of energy from collective into
intrinsic single-particle motion: two-body collisions and ``one-body
friction''. The latter is caused by the moving walls of the
self-consistent nuclear mean field. The role played by these two
dissipation mechanisms in fission and heavy-ion reactions is not yet 
completely understood. In a pioneering article that appeared in 1976
Davies, Sierk and Nix \cite{DSN76} calculated the effect of
viscosity on the dynamics of fission. Assuming that friction is caused by
two-body collisions they extracted a viscosity coefficient $\mu = 0.015$
Tera Poise from a comparison of
theoretical and experimental values for the kinetic
energies of fission fragments. The corresponding time delay for the
nuclear motion from the saddle to the scission point was found to be of
order $\Delta t =1 \times 10^{-21}$ s. However, in one-body dissipation
models the time delay is an order of magnitude larger.
Several experimental techniques are sensitive to the energy dissipation
in nuclear fission. At high excitation energy, the multiplicity of
pre-scission neutrons \cite{Ga87} or photons \cite{Ho95} depends on the
dissipation strength. At low excitation energy, the process of
prompt muon-induced fission \cite{MO80} provides a suitable ``clock''.
This process will be discussed here.

After muons have been captured into high-lying single particle states
they form an excited muonic
atom. Inner shell transitions may proceed without photon emission by
inverse internal conversion, i.e. the muonic excitation energy is
transferred to the nucleus. In actinides, the $2p \rightarrow 1s$ and the
$3d \rightarrow 1s$ muonic transitions result in excitation of the nuclear
giant dipole and giant quadrupole resonance, respectively, which act as
doorway states for fission. The nuclear excitation energy is typically
between 6.5 and 10 MeV. Most importantly, the muon is still available
following these atomic transitions 
(in the ground state of the muonic atom) and can be utilized to probe
the fission dynamics. Eventually though, the muon will 
disappear as a result of the weak interaction (nuclear capture by
one of the fission fragments). However, the nuclear capture occurs
on a time scale of order $10^{-7}$ s which is many orders
of magnitude larger than the time scale of fission.

The prompt muon-induced fission process is most easily understood via a
``correlation diagram'', i.e. one plots the single-particle energies of the
transient muonic molecule as a function of the internuclear distance
\cite{OU93}. If there is a large amount of friction during the motion
from the outer fission barrier to the scission point the muon will
remain in the lowest molecular energy
level $1s\sigma$ and emerge in the $1s$ bound
state of the {\it heavy} fission fragment. If, on the other hand, friction is
small and hence the nuclear collective motion is relatively
fast there is a nonvanishing probability
that the muon may be transferred to higher-lying molecular orbitals, e.g.
the $2p\sigma$ level, from where it will end up attached to the {\it light}
fission fragment. Therefore, theoretical studies of the muon-attachment
probability to the light fission fragment, $P_L$, in combination with
experimental data can be utilized to analyze the dynamics of fission,
and nuclear energy dissipation in particular.

\section{Theoretical Developments}

Because the nuclear excitation energy in muon-induced fission exceeds
the fission barrier height it is justified to treat the
fission dynamics classically (no barrier tunneling). For simplicity,
we describe the fission path by one collective coordinate $R$;
the classical collective nuclear energy has the form 
\begin{equation}
E_{\rm nuc} = \frac{1}{2} B(R) \dot R^2 + V_{\rm fis}(R) + E_\mu (R). 
\label{ecoll}
\end{equation}
We utilize a coordinate dependent mass parameter \cite{OU93} and an empirical
double-humped fission potential $V_{\rm fis}(R)$ \cite{Ba74} which
is smoothly joined with the Coulomb potential of the fission fragments at
large $R$. The last term in Eq. (\ref{ecoll}) denotes the instantaneous muonic
binding energy which depends on the fission coordinate; this term will
be defined later. 

To account for the nuclear energy dissipation between the outer fission barrier
and the scission point, we introduce a friction force which depends 
linearly on the velocity. In this case, the dissipation function $D$ is a simple
quadratic form in the velocity
\begin{equation}
\dot E_{\rm nuc}(t) = -2D = -f \dot R^2 (t) \label{frict}.
\end{equation}
The adjustable friction parameter $f$ determines the dissipated energy; it
is the only unknown quantity in the theory.

For the dynamical description of the muonic wavefunction during prompt
fission, the electromagnetic coupling between muon and nucleus $(-e \gamma
_\mu A^\mu)$ is dominant; the weak interaction is negligible. Because
of the nonrelativistic motion of the fission fragments the
electromagnetic interaction is dominated by the Coulomb interaction
\begin{equation}
A^0({\bf r},t) = \int d^3r' \frac{\rho_{\rm nuc}( {\bf r'},t)}
{| {\bf r} - {\bf r'} |}.    \label{vcoul}
\end{equation}
The muonic binding energy in the ground state of an actinide muonic atom
amounts to 12 percent of the muonic rest mass; hence nonrelativistic
calculations, while qualitatively correct, are limited in accuracy. Several
theory groups have demonstrated the feasibility of such calculations
\cite{MO80,MW81,Ka97} which are based on the time-dependent Schr\"odinger equation
\begin{equation}
[ -\frac{\hbar^2}{2m} \nabla^2 -e A^0 ({\bf r},t) ] \ 
\psi ({\bf r},t) = i \hbar \frac{\partial}{\partial t} \ \psi ({\bf r},t) .
\end{equation}
Recently, we have developed a numerical algorithm to solve the relativistic
problem on a three-dimensional Cartesian mesh \cite{OU92,OU93}.
The time-dependent Dirac equation for
the muonic spinor wave function in the Coulomb field of the fissioning
nucleus has the form
\begin{equation}
H_{\rm D}(t) \ \psi ({\bf r},t) = i \hbar \frac \partial
{\partial t} \ \psi ({\bf r},t) \label{tdirac},
\end{equation}
where the Dirac Hamiltonian is given by
\begin{equation}
H_{\rm D}(t) =  -i \hbar c {\bf \alpha} \cdot \nabla + \beta mc^2
-e A^0 ({\bf r},t). \label{hdirac}
\end{equation}
Our main task is the solution of the Dirac
equation for the muon in the presence of a time-dependent external Coulomb
field $A^0({\bf r},t)$ which is generated by the fission
fragments in motion.
Note the coupling between the fission dynamics, Eq. (\ref{ecoll}), and
the muon dynamics, Eq. (\ref{tdirac}), via the
instantaneous muonic binding energy
\begin{equation}
E_\mu (R(t)) = \langle \psi ({\bf r},t) \mid H_{\rm D}(t) \mid
\psi ({\bf r},t) \rangle
\end{equation}
which depends on the fission coordinate; the presence of this
term increases the effective fission barrier height.

\section{Lattice Representation: Basis-Spline Expansion}

For the numerical solution of the time-dependent Dirac equation (\ref{tdirac})
it is convenient to introduce dimensionless space and time coordinates

$$
{\bf x} = {\bf r} / \lambdabar_c \ \ \ \ \lambdabar_c = \hbar /(m_\mu c)=1.87 fm
$$
\begin{equation}
\tau = t / \tau _c \ \ \ \ \tau _c= \lambdabar_c / c = 6.23 \times 10^{-24} s
\label{comptim}
\end{equation}
where $\lambdabar_c$ denotes the reduced Compton wavelength of the muon and
$\tau _c$ the reduced Compton time. For the lattice representation of the
Dirac Hamiltonian and spinor wave functions we introduce
a 3-dimensional rectangular box with a uniform lattice spacing $\Delta x$.
The lattice points are labeled $( x_\alpha, y_\beta, z_\gamma)$.

Our numerical algorithm is the Basis-Spline collocation method \cite{WO95}.
Basis-Spline functions $B_i^M(x)$ are piecewise-continuous polynomials
of order $(M-1)$. These may be thought of as generalizations of the 
well-known ``finite elements'' which are a B-Splines with $M=2$.
To illustrate the method let us consider
a wave function which depends on one space coordinate $x$;
we represent the wave function on a finite spatial interval
as a linear superposition of B-Spline functions 
\begin{equation}
\psi(x_\alpha) = \sum_{i=1}^{N} B^M_i(x_\alpha) c^i .
\label{psialpha}
\end{equation}
In the Basis-Spline collocation method, local operators such
as the EM potential $A^0$ in Eq. (\ref{hdirac})
become diagonal matrices of their values at the grid points
(collocation points), i.e.\ $V(x) \rightarrow V_\alpha=V(x_\alpha)$.
The matrix representation of derivative operators is
more involved \cite{WO95}. For example, the first-derivative
operator of the Dirac equation has the following
matrix representation on the lattice
\begin{equation}
D_\alpha^\beta
\equiv  \sum_{i=1}^{N} B'_{\alpha i} B^{i \beta}\;,
\label{1der}
\end{equation}
where $B'_{\alpha i} = [dB_i^M(x) / dx] |_{x=x_\alpha}$.
Furthermore, we use the shorthand notation
$B_{\beta i}=B^M_i(x_\beta)$
for the B-spline function evaluated at the collocation point $x_\beta$,
and the inverse of this matrix is denoted by $B^{i \beta} =
[B^{-1}]_{\beta i}$.
Because of the presence of this inverse, the operator
$D_\alpha^\beta$ will have a nonsparse matrix representation.
In the present calculations we employ B-Splines of
order $M=5$. Eq. (\ref{psialpha}) can readily be generalized to three
space dimensions; in this case the four Dirac
spinor components $\psi ^{(p)}, p=( 1,\cdot \cdot \cdot,4)$
are expanded in terms of a product of Basis-Spline functions
\begin{equation}
\psi ^{(p)}( x_\alpha ,y_\beta ,z_\gamma ,t) = 
\sum\limits_{i,j,k}B^M_i(x_\alpha )B^M_j(y_\beta )B^M_k(z_\gamma )
c_{(p)}^{ijk}(t) ,
\end{equation}
i.e. the lattice representation of the spinor wave function
is a vector with $N = 4 \times N_x \times N_y \times N_z$ 
complex components. Hence,
it is impossible to store $H_{\rm D}$ in memory because this would
require the storage of  $N^2$ complex double-precision numbers.
We must therefore resort to iterative methods for the solution of the matrix
equation which do not require the storage of $H_{\rm D}$.

We solve the time-dependent Dirac equation in two steps: first, we solve
the static Coulomb problem at time $t=0$, i.e. the muon bound to an
actinide nucleus. This problem is solved by the damped relaxation
method \cite{OU93}. The second part of our numerical procedure is
the solution of the time-dependent Dirac equation (\ref{tdirac})
by a Taylor-expansion of the propagator for an infinitesimal time step
$\Delta t$. Details may be found in ref. \cite{OU93}.

\section{Discussion of Numerical Results}
In the following we present results for prompt fission of $^{237}_{\ 93}$Np
induced by the $3d \rightarrow 1s$ muonic transition $(9.5 {\rm MeV})$.
All results reported here are for a 3-D Cartesian lattice of size
$L_x = L_y = 67$ fm and $L_z = 146$ fm with $N_x \times N_y \times N_z =
25 \times 25 \times 53$ lattice points with a uniform lattice spacing
$\Delta x = 1.5 \lambdabar_c = 2.8 fm$. Depending on the value of the
friction coefficient, we utilize between $1,200 - 1,900$ time steps with
a step size $\Delta t = 1.5 \tau_c = 9.3 \times 10^{-24}$ s.
Typical production runs take about 11 hours of CPU time on a CRAY
supercomputer or about 54 hours on an IBM RS/6000 workstation.

\begin{figure}[t]
\vspace*{-1.0cm}
                 \epsfxsize=12cm \epsfbox{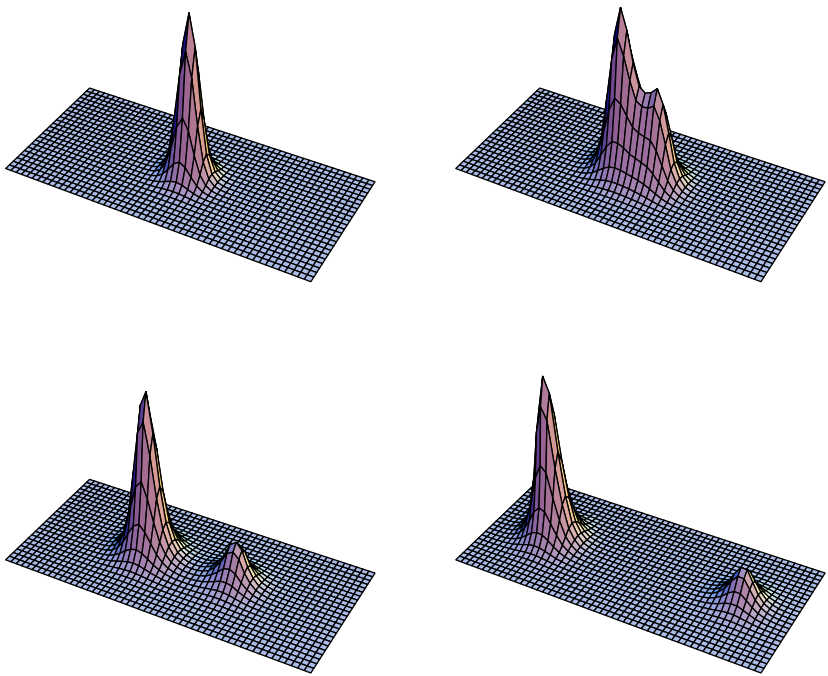}
\vspace*{-0.5cm}
\caption[]{
Prompt muon-induced fission of $^{237}_{\ 93}$Np for a fission fragment mass
asymmetry $\xi = A_H / A_L = 1.10$ at $E^* = 9.5$ MeV. Shown is the muon 
position probability density at four different times during fission: 
$t = 0$, $6.5 \times 10^{-21}$ s, $8.4 \times 10^{-21}$ s, $1.1 \times
10^{-20}$ s. Zero friction ($f=0$) has been assumed.
}
\label{fig1}
\end{figure}
 
Fig.\ \ref{fig1} shows the time-development of the muon position probability
density during fission at a fragment mass asymmetry $\xi = A_H / A_L = 1.10$.
As expected, the muon sticks predominantly to the heavy fragment, but for this
small mass asymmetry the muon attachment probability to the light fission
fragment, $P_L$, is rather large (20 percent).

One might ask whether the muon will always remain bound during fission;
what is the probability for ionization? To investigate this question we
have plotted the muon position probability density on a logarithmic scale.

\begin{figure}[htb]
\vspace*{0.2cm}
				\epsfxsize=12cm \epsfbox{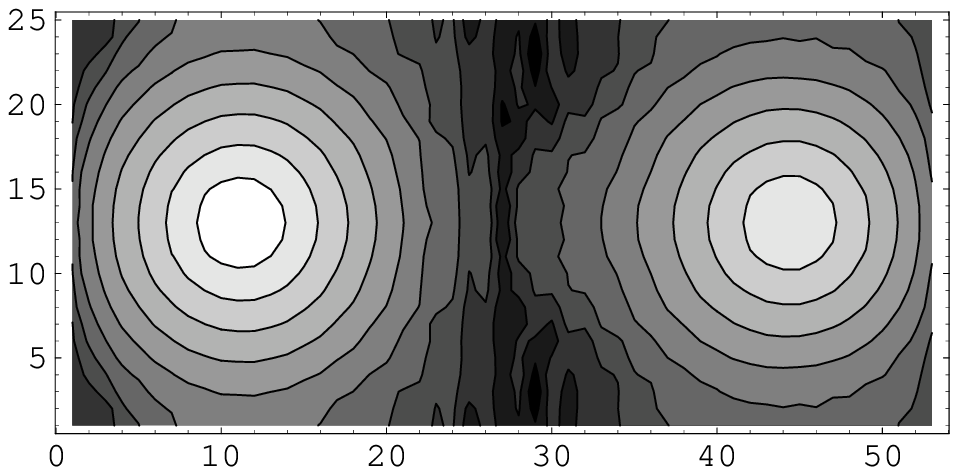}
\vspace*{0.2cm}
\caption[]{
Contour plot of the logarithm of the muon probability density at $ t =  
1.1 \times 10^{-20}$ s shows no evidence of muon ionization.
}
\label{fig2}
\end{figure}

In coordinate space, any appreciable muon ionization would show up as a
``probability cloud'' that is separating from the fission fragments and
moving towards the boundaries of the lattice. Fig.\ \ref{fig2} shows no
evidence for such an event in our numerical calculations. Hence, we conclude
that the probability for muon ionization $P_{\rm ion}$ is
substantially smaller than the muon attachment probability to the light
fission fragment which is always clearly visible in our logarithmic plots,
even at large mass asymmetry. From this we estimate that $P_{\rm ion}
< 10^{-4}$.

\begin{figure}[htb]
\vspace*{0.2cm}
                 \epsfxsize=12cm \epsfbox{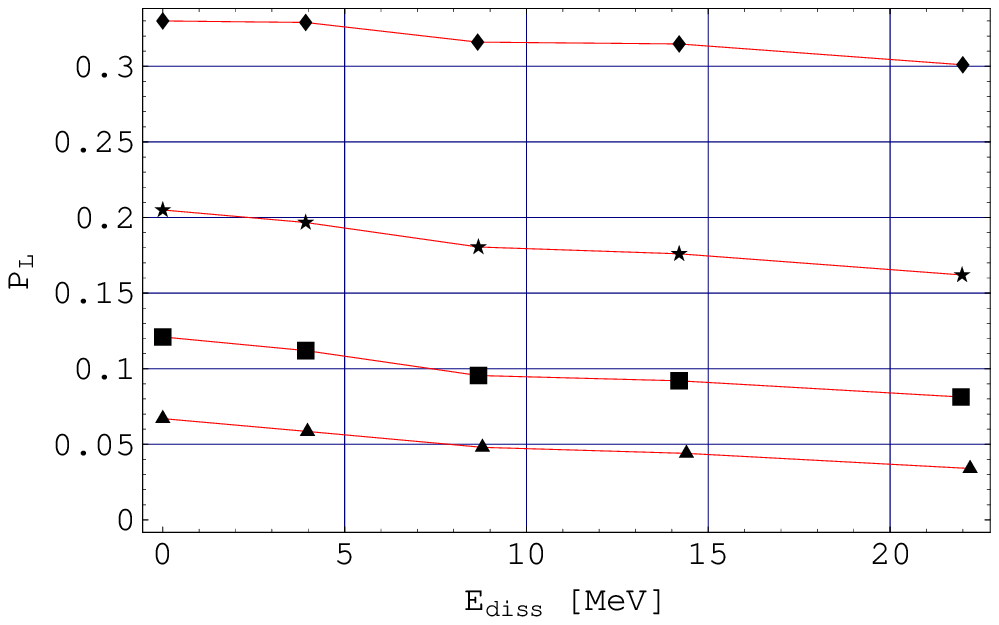}
\vspace*{0.2cm}
\caption[]{
Muon attachment probability to the light fission fragment as
function of nuclear energy dissipation for $^{237}_{\ 93}$Np. Results are
shown for fragment mass asymmetries $\xi=1.05$ (upper curve), $1.10,\ 1.15$,
and $1.20$ (lower curve).
}
\label{fig3}
\end{figure}
 
Fig. \ref{fig3} shows that $P_L$ depends strongly on the fission fragment
mass asymmetry. This is easily understood: for equal fragments we obviously
obtain $P_L=0.5$, and for large mass asymmetry it is energetically favorable
for the muon to be bound to the heavy fragment, hence $P_L$ will be small.
In Fig.\ \ref{fig3} we also examine the dependence of $P_L$ on the dissipated
nuclear energy, $E_{\rm diss}$, during fission. In our model, friction
takes place between the outer fission barrier and the scission point.
When the dissipated energy is computed from equation (\ref{frict})
we find an almost linear dependence of the muon attachment probability on
$E_{\rm diss}$; unfortunately, this dependence is rather weak.

We would like to point out that the theoretical values for $P_L$ obtained
in this work are smaller than those reported in our earlier calculations
\cite{OU92,OU93}. There are two reasons for this: (a) the size of
the lattice and (b) the lattice representation of the first derivative
operator in the Dirac equation. Because of constraints in the amount
of computer time available to us we utilized a smaller cubic lattice
in our prior calculations \cite{OU93} with
$N_x \times N_y \times N_z = 29^3$ lattice points. 
More recently, we were able to increase the size of the lattice
substantially, in particular in fission ($z$-) direction (see above).
In Fig. 2 of ref. \cite{OU98} we have demonstrated the convergence of
our results for the muon attachment probability in terms of the lattice
size and lattice spacing. Another reason for the difference between the
current and prior results is the lattice representation of the first
derivative operator, Eq. (\ref{1der}), in the Dirac equation.
In ref. \cite{OU92,OU93} we utilized a combination of forward and backward
derivatives for the upper and lower spinor wave function components; after
extensive testing of Coulomb potential model problems with known analytical
solutions we have found that the symmetric derivative operator
provides a more faithful lattice representation. The results reported
here and in ref. \cite{OU98} have been obtained utilizing the symmetric
derivative prescription.

\section{Comparison of Theory with Experiment}
There are only a few experimental data available for comparison.
Schr\"oder {\it et al.} \cite{SW79} measured for the first time
mean lifetimes of muons bound to fission fragments of several
actinide nuclei. The muon decays from the K-shell of the muonic atom
through various weak interaction processes at a characteristic rate 
$\lambda = \lambda_0 + \lambda_c$, where $\lambda_0 = 
(2.2\times 10^{-6}s)^{-1}$ is the free leptonic decay rate for 
the decay process $\mu^- \rightarrow e^- + \bar{\nu_e} + \nu_{\mu}$ and
$\lambda_c$ denotes the nuclear capture rate; $\lambda_c$ depends upon the
charge and mass of the fission fragment. From the observed lifetime $\tau _\mu
=1.30\times 10^{-7}s$ Schr\"oder {\it et al.} estimated an upper limit
for the muon attachment probability $P_L \le 0.1$. It must be 
emphasized that this number represents an integral over the whole fission
mass distribution and, hence, cannot be directly compared to the
numbers given in Fig. \ref{fig3}.

The most complete experiments have been carried out by Risse {\it et al.}
\cite{RB91} at the Paul Scherrer Institute (PSI) in Switzerland. The basic
experimental approach is to place a fission
chamber inside an electron spectrometer. The
incident muons are detected by a scintillation counter. An event
is defined by a $(\mu^-, f_1 f_2 e^-)$ coincidence where the fission
fragments are observed in prompt and the muon decay electrons in delayed
coincidence with respect to the incident muon. The magnetic field of the
electron spectrometer allows for a reconstruction of the
electron trajectories. Thus, it is possible to determine whether the muon
decay electrons originate from the heavy or the light fission fragment.

For several mass bins of the light fission fragment,
muon attachment probabilities $P_L$ have been measured; the experimental
data are given in Table \ref{exptheo}. It should be emphasized that the
mass bins are relatively broad. Because the theoretical values for $P_L$ 
depend strongly on the mass asymmetry it is not justified to assume that
$P_L$ remains constant within each experimental mass bin.
Instead, to allow for a comparison between theory and experiment,
we have to multiply the theoretical $P_L$ values in Fig.\ \ref{fig3}
with a weighting factor that accounts for the measured relative mass
distribution \cite{RB91} of the prompt fission events within this mass bin.
We subsequently integrate the results over the sizes of the experimental
mass bins.
Due to the relatively low excitation energy in muon-induced fission,
the fission mass distribution exhibits a maximum at $\xi = A_H / A_L = 1.4$
and falls off rather steeply for values larger or smaller than the maximum.
This means that the large values of $P_L \approx 0.5$ at or near
fission fragment symmetry $\xi=1.0$ will be strongly suppressed. 
The resulting theoretical values for $P_L$ are given in the last column of
Table \ref{exptheo}. It is apparent that our theory agrees rather well 
with experiment. Because of the size of the error bars in the experiment
and because of the weak dependence of the theoretical values of $P_L$ on
the dissipated energy, it is not possible to extract very precise information
about the amount of energy dissipation.

\begin{table}[!t]
\caption{Muon-attachment probabilities to the light fission fragment,
$P_L$, for $^{237}{\rm Np}(\mu^-,f)$. Exp. data are taken from ref.
\cite{RB91}.
\label{exptheo}}
\vspace{0.2cm}
\begin{center}
\footnotesize
\begin{tabular}{|c|c|c|c|}
\hline
{mass bin $A_L$} &\raisebox{0pt}[13pt][7pt]{mass asymmetry} &
\raisebox{0pt}[13pt][7pt]{$P_L$(exp)} &{$P_L$(theo)}\\
\hline
 & & & \\
$118.5 \rightarrow 111.5$   & $1.000 \rightarrow 1.126$ 
& $(25.5 \pm 8.5) \times 10^{-2}$
& $26.0 \times 10^{-2}, E_{\rm diss}=0 {\rm MeV}$ \\
    &    &    & $22.3 \times 10^{-2}, E_{\rm diss}=22 {\rm MeV}$ \\
 & & & \\
\hline
 & & & \\
$111.5 \rightarrow 104.5$   & $1.126 \rightarrow 1.268$ 
& $(9.7 \pm 2.6) \times 10^{-2}$
& $6.62 \times 10^{-2}, E_{\rm diss}=0 {\rm MeV}$ \\
    &    &    & $3.51 \times 10^{-2}, E_{\rm diss}=22 {\rm MeV}$ \\
 & & & \\
\hline
\end{tabular}
\end{center}
\end{table}

From a comparison of our theoretical result for the mass bin
$A_L = 118.5 \rightarrow 111.5$ with the measured data we extract
a dissipated energy of order $10$ MeV for $^{237}$Np while the second
mass bin $A_L = 111.5 \rightarrow 104.5$ is more compatible with zero
dissipation energy. We place a higher confidence on the theoretical
results for the first mass bin because the probabilities $P_L$ are
substantially larger and hence numerically more reliable. We like
to point out that our theoretical value $E_{\rm diss}=10$ MeV is
compatible with results from other low-energy fission
measurements that are based on the odd-even effect in the charge yields
of fission fragments \cite{Wa91}. In addition to $^{237}$Np
we have also studied muon-induced fission of $^{238}$U; the results
for muon attachment are very similar \cite{OU98}.

\section{Conclusions}
We have studied the dynamics of a muon bound to a fissioning
actinide nucleus by solving the time-dependent Dirac equation for the
muonic spinor wavefunction; the fission dynamics is described classically.
The theory predicts a strong mass asymmetry dependence of the muon
attachment probability $P_L$ to the light fission fragment; this feature
is in agreement with experimental data. Our calculations show no evidence
for muon ionization during fission. The theory also predicts
a (relatively weak) dependence of $P_L$ on the dissipated energy. By
comparing our theoretical results to the experimental data of
ref. \cite{RB91} we extract a dissipated energy of about $10$ MeV for
$^{237}$Np (see Table 1). Using the dissipation function defined in
Eq. (\ref{frict}), this value corresponds to a fission time delay from
saddle to scission of order $2 \times 10^{-21}$ s.

\section*{Acknowledgements}
This research project was sponsored by the U.S. Department of Energy under
contract No. DE-FG02-96ER40975 with Vanderbilt University. For several years,
I have benefitted from fruitful discussions with my collaborators, in particular
with J.A. Maruhn, the late C. Bottcher, M.R. Strayer, P.G. Reinhard, A.S. Umar
and J.C. Wells. Some of the numerical calculations were carried out on CRAY
supercomputers at NERSC, Berkeley. I also acknowledge travel support to
Germany from the NATO Collaborative Research Grants Program.

\vfill\eject
\end{document}